\documentclass[
reprint,
 amsmath,amssymb,
 aps,
floatfix,
]{revtex4-2}
\usepackage{lineno}
\usepackage{comment}
\usepackage{graphicx}
\usepackage{dcolumn}
\usepackage{bm}
\usepackage{color}
\usepackage{xcolor}
\usepackage{mathtools}

\begin{document}

\preprint{APS/123-QED}
\title{Pair momentum dependence of tilted source in heavy ion collisions}

\author{Yevheniia Khyzhniak}
\author{Michael Annan Lisa}
\affiliation{
 Physics Department, The Ohio State University, Columbus, Ohio 43210, USA
}

\date{\today}

\begin{abstract}

In non-central heavy-ion collisions, the particle-emitting source can be tilted away from the beam direction, an effect that becomes particularly significant at collision energies of a few GeV and lower. This phenomenon, manifest itself in many observables such as directed flow, polarization, and vorticity, is therefore important to investigate. In this paper, we study the consistency between the tilt extracted directly from the freeze-out distribution of pions and the tilt parameter obtained using the azimuthally sensitive femtoscopy (asHBT) method. Using the UrQMD model, we demonstrate a strong dependence of the tilt parameter extracted with asHBT on the momentum of the particle pair. Considering the experimental challenges in accessing low particle momenta—where the tilt parameter extracted with asHBT closely matches the tilt of the freeze-out distribution of pions—we propose an exponential extrapolation method to obtain the tilt of the entire freeze-out distribution. This approach aims to enhance the accuracy of experimental measurements of tilt in non-central heavy-ion collisions.

\end{abstract}
\maketitle

\section{Introduction}

Heavy-ion collisions provide unique ways to study the properties of hot and dense matter under extreme conditions. Over the years, the initial state and its three-dimensional structure have become significant topics of interest in the community~\cite{Arslandok:2023utm, Achenbach:2023pba}. These properties cannot be accessed and measured directly in the experiment. Instead, one can study momentum anisotropy 
and correlations that appear in the initial stages of the collisions but propagate to the kinetic freeze-out.

To describe  the fluid-dynamic properties of the medium formed in a collision, one uses the concept of anisotropic flow, which is characterized by Fourier coefficients that quantify different patterns of collective particle emission~\cite{Poskanzer:1998yz}. The coefficient $v_{1}$, known as directed flow, corresponds to the side-ward deflection of particles. The coefficient $v_{2}$, or elliptic flow, describes the elliptical shape of the particle distribution in the transverse plane. Higher-order harmonics, such as $v_{3}$ and beyond, capture more complex flow patterns and higher-order anisotropies in particle emission.

\begin{figure}
    \centering
    \includegraphics[width=1.0\linewidth]{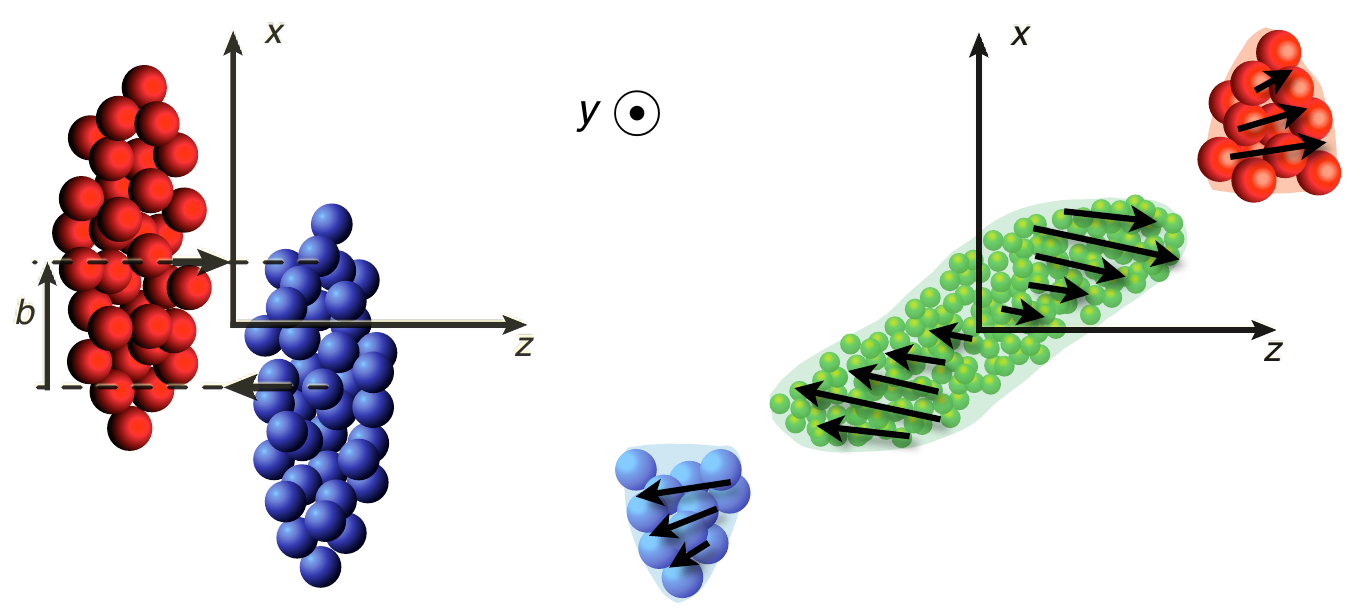}
    \caption{Schematic view of the tilted emission source at the freeze-out time which appear in the non-central collisions. }
    \label{fig:collision}
\end{figure}

While $v_{2}$ has been successfully described in various studies~\cite{Schenke:2020mbo,Gale:2012rq}, directed flow remains an area of active research, particularly concerning its slope, denoted as $dv_{1}/dy$~\cite{STAR:2014clz,Jiang:2023fad,Jing:2022ehp,Du:2022yok,Bozek:2022svy}. The presence of directed flow is attributed to the overall tilted emitting particle source, which results from forward-backward asymmetry of the initial geometry of the colliding nuclei. Therefore, to accurately describe the azimuthal distribution of the emitting particles, it is essential to measure the tilt of the emission source. This tilt effect, arising from the initial colliding geometry, is illustrated in the schematic picture of the collision in Fig~\ref{fig:collision}. On the left side of the picture, two nuclei can be seen before the collision, and on the right side, a tilted emitting particle source created during the collision, along with spectators, is visible. The arrows indicate the direction of motion of particles, which will be reflected in the directed flow and other observables.

It has been shown that in order to recover the tilt of of the emitting source, one can use the method of azimuthally-sensitive femtoscopy (or as people often refer to it - asHBT), which probes the spatial and temporal scales of particle emission using momentum correlations relative to the reaction plane (plane defined by the impact parameter and collision direction)~\cite{Retiere:2003kf}. In the 1970s, G.I. Kopylov and M.I. Podgoretsky demonstrated the possibility of using momentum correlations (quantum-statistical) to study the spatiotemporal characteristics of particle formation processes in particle and nuclear collisions~\cite{Kopylov:1972qw,Kopylov:1975rp,Kopylov:1974th,Podgoretsky:1989bp}.

They specifically proposed studying the interference effect using the correlation function, as well as employing the event-mixing method to construct reference distributions that do not contain quantum-statistical correlations. Subsequent theoretical work, combined with experimental measurements, transformed the method of correlation femtoscopy into a precise tool for investigating particle production and final state interactions~\cite{Lednicky:1995vk,Lednicky:2005tb,Voloshin:1997jh,Lednicky:1981su}.

As the research continued, it became evident that the correlations between identical particles emitted by highly excited nuclei are influenced not only by the system's spatial structure but also by its lifetime~\cite{Kopylov:1972qw,Shuryak:1972kq}. Further insights revealed that the momentum dependence of particle pair correlations observed in relativistic heavy-ion collisions provides valuable information about the dynamics of the collision process~\cite{Pratt:1984su,Pratt:1986ev}.

\begin{figure}
    \centering
    \includegraphics[width=1.0\linewidth]{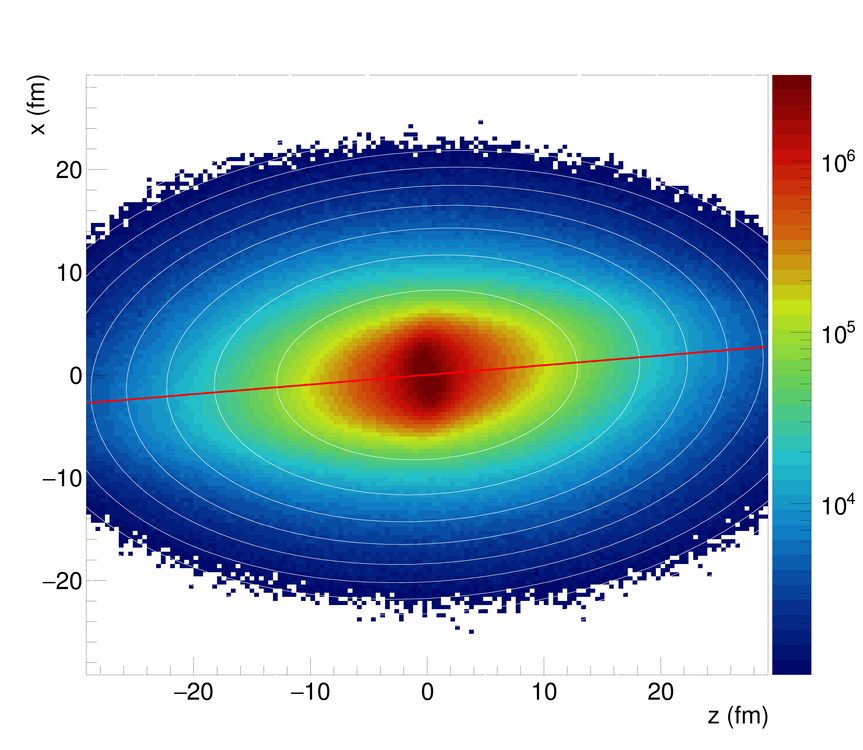}
    \caption{Two-dimensional projection of the three-dimensional distribution of the pion position in space obtained in the Au+Au collisions at 7.7 GeV using UrQMD model. White lines represent the two-dimensional projection of the three-dimensional fit performed according to Equation~\ref{eq:freeze-out}. The red line represents the tilt of this distribution obtained from the three-dimensional fit.}
    \label{fig:freeze-out}
\end{figure}

In a Ref.~\cite{Mount:2010ey}, the particle emission source was analyzed using a minimally generalized source description: a tilted Gaussian ellipsoid as in Equation~\ref{eq:freeze-out}. Later, it was shown that the situation is not so straightforward and that the tilt of the source may depend on the radius of the fit sphere over which the distribution is being considered~\cite{Graef:2013wta}. However, ultimately, as one goes to larger radius of the sphere the tilt values seems to level off~\cite{Graef:2013wta}. The example of the three-dimensional freeze-out distribution projected onto $``x-z"$ plane and it's fit according to the Equation~\ref{eq:freeze-out} is shown in Figure~\ref{fig:freeze-out}. The red line represents $\theta$ angle which is extracted from the three-dimensional fit according to:

\begin{align}
 f(x,y,z) &= N\exp \left( -\frac{x^{'2}}{2\sigma_{x^{'}}}-\frac{y^{2}}{2\sigma_{y^{}}} -\frac{z^{'2}}{2\sigma_{z^{'}}}      \right) \label{eq:freeze-out} \\
x^{'} &=(x\cos\theta - z\sin\theta)^2  \label{eq:rot1}\\
z^{'} &=(x\sin\theta + z\cos\theta)^2  \label{eq:rot2}
\end{align}
where N is the normalization; $x, y$ and $z$ are spatial coordinates of the pions freeze-out; $\theta$ is the angle of rotation and $x^{'}$ and $z^{'}$ rotated coordinates on the angle $\theta$ as it shown in the Equations~\ref{eq:rot1} and~\ref{eq:rot2}.

Unfortunately, the same measurement cannot be performed in an experiment. One cannot directly gain access to the information of the spatial coordinates where particles were created or froze out. In Ref.~\cite{Wiedemann:1997cr} and later on in Ref.~\cite{Lisa:2000ip} the method of azimuthally sensitive femtoscopy, which can be used to experimentally measure the tilt angle of the emission source of certain particles was proposed. The same year, the first experimental measurement by the E895 collaboration was performed~\cite{E895:2000opr}. The results showed that the tilt of the emission source can be quite large (up to $\thicksim$ 50 degrees in this study) and also strongly dependent on energy of the collision~\cite{E895:2000opr}. This study confirmed that accounting for the source tilt, especially at low energies, is essential, and that using boost invariant modeling of heavy-ion collisions will inevitably lead to an incorrect description of the system's evolution. 
Over the years, boost-invariant models that ignore the three-dimensional structure of the initial state have been quite successful in describing the transverse profile of initial energy density distributions. This includes the most widely used 2D initial state models such as Mc-Glauber~\cite{Schenke:2011bn}, MC-KLN~\cite{Hirano:2005xf}, EKRT~\cite{Niemi:2015qia}, TRENTO~\cite{Moreland:2014oya}, and IP-Glasma~\cite{Schenke:2012wb}. However, going forward, to capture the tilted geometry, one needs to go beyond these approximations and develop more sophisticated models that account for the three-dimensional structure of heavy-ion collisions.

\section{Azimuthally sensitive femtoscopic correlations}

To study a realistic picture of the particle emission source emerging in heavy ion collisions, the Ultrarelativistic Quantum Molecular Dynamics (UrQMD 3.4) model~\cite{Bass:1998ca,Bleicher:1999xi} was used in cascade mode. In this work, we focus on the pion emission source. Pions with transverse momentum less than 0.8 GeV/c and pseudorapidity $|\eta| < 1$ were used to match future experimental studies. We considered two collision energies: 7.7 GeV and 27 GeV.

Femtoscopic analysis requires the presence of Bose-Einstein correlations between identical particles in the model. However, incorporating these correlations is challenging because they arise at the amplitude level, while most commonly used models, such as Monte Carlo event generators, are formulated at the probability level. Consequently, Bose-Einstein and other correlations at the multi-particle amplitude level in the initial conditions are not accounted for in the dynamical evolution of collisions in these models. This results in a loss of quantum mechanical correlations, which are not present at the output level of such event generators. Therefore, it is necessary to restore these correlations in the simulated particle distribution by applying weights. By using Bose-Einstein weights, one can transform the discrete phase-space coordinates $(r_i, t_i, p_i)$ into a correlated distribution of particles, encoded in distributions such as $C(q, K)$.

\begin{figure}
    \centering
    \includegraphics[width=0.86\linewidth]{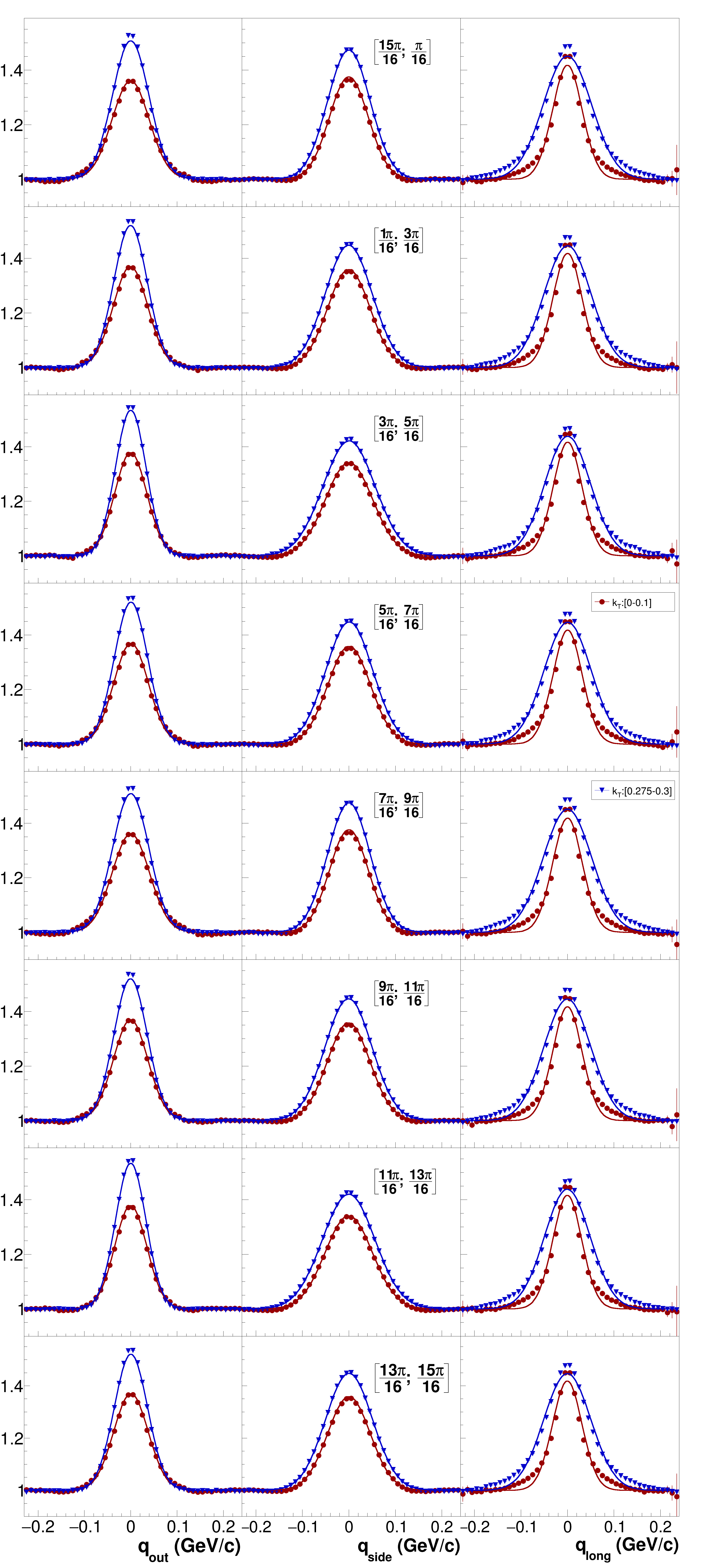}
    \caption{One-dimensional projections of the three-dimensional correlation functions constructed for different azimuthal angles of the pion pairs in 30-50\% centrality range of the Au+Au collisions at 7.7 GeV. Different colors and markers correspond to different transverse momentum of the pion pair: red circles - from 0 to 0.1 GeV/c, blue triangles - from 0.275 to 0.3 GeV/c. The lines correspond to the projections of the three-dimensional fit of the correlation functions according to Equation~\ref{eq:cfFit}.}
    \label{fig:1DCF}
\end{figure}

\begin{figure}
    \centering
    \includegraphics[width=0.86\linewidth]{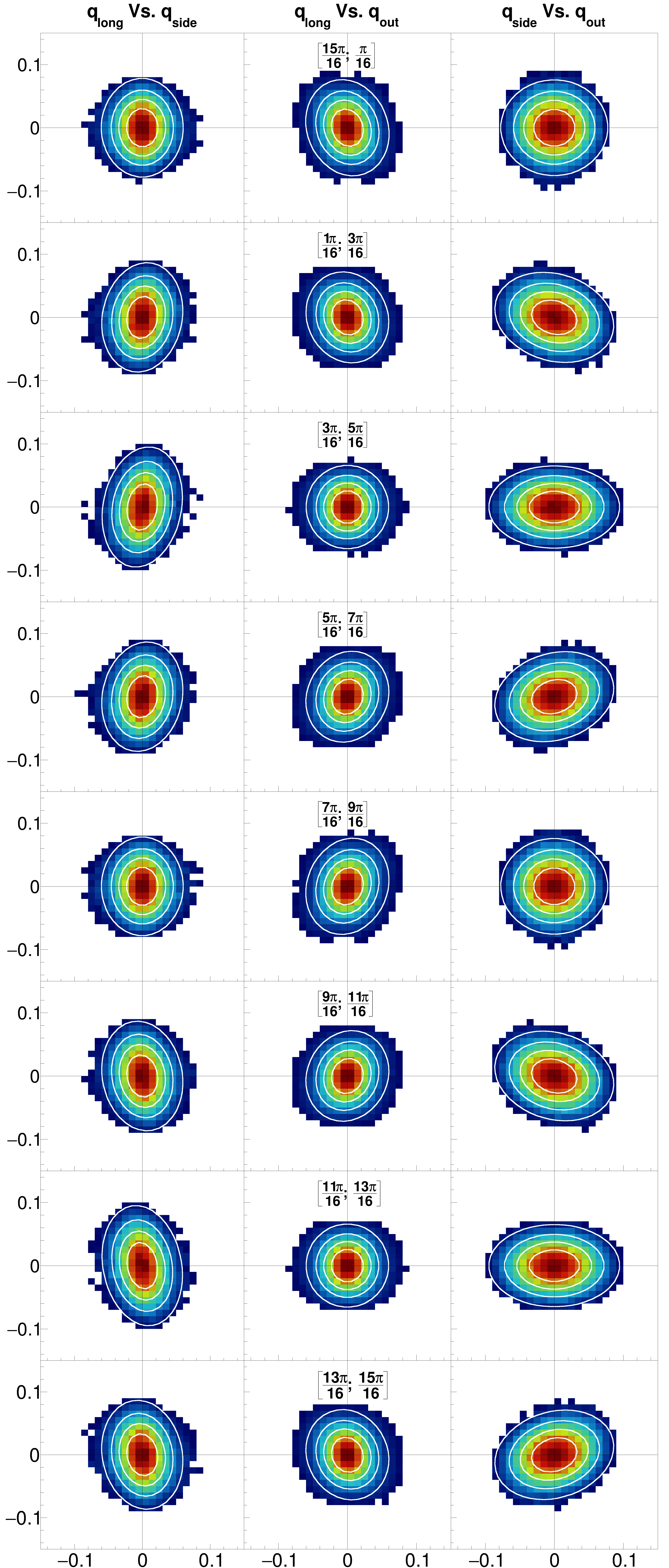}
    \caption{Two-dimensional projections of the three-dimensional correlation functions constructed for different azimuthal angles of the pion pairs with transverse momentum from 0 to 0.1 GeV/c in 30-50\% centrality range of the Au+Au collisions at 7.7 GeV. White contours correspond to the two-dimensional projections of the three-dimensional fit of the correlation functions according to the Equation~\ref{eq:cfFit}.}
    \label{fig:2DCF}
\end{figure}

According to Ref.~\cite{Wiedemann:1999qn}, quantum-statistical correlations between identical particles can be added to the model using:

\begin{equation}
C(\textbf{q},\textbf{K}) = 1 + \int d^4x\cos(q\cdot x) \,d(x,K) 
\label{eq:theorOsc}
\end{equation}
where $C(\textbf{q},\textbf{K})$ is the two-particle correlation function for pairs of identical bosons which characterize homogenity region~\cite{Akkelin:1995gh}; $q = p_{1}-p_{2}$ is the relative pair momentum; $K=\frac{p_{1}+p_{2}}{2}$ is the average pair momentum; $x = x_{1}-x_{2}$ is the separation between particles in the pair.  Examples of the correlation functions constructed at 30-50\% collision centrality can be seen in Figures~\ref{fig:1DCF} and~\ref{fig:2DCF}. Figure~\ref{fig:1DCF} shows one-dimensional projections of the three-dimensional correlation function onto the  $q_{out}$, $q_{side}$ and $q_{long}$ axes, corresponding to axes in the Bertsch-Pratt parameterization~\cite{Pratt:1986cc,Bertsch:1988db}. When projecting onto one axis, the other two were kept within the range $|q_{other}|< 50$~MeV/c. The projections are shown for two different ranges on transverse momentum of the pion pair. In this figure, the red circles and lines show the projections of the correlation function constructed in the transverse momentum range of the pion pairs from 0 to 100 MeV/c, while the blue triangles and lines show the same but in the range from 300 to 350 MeV/c.

\begin{figure}
    \centering
    \includegraphics[width=1.0\linewidth]{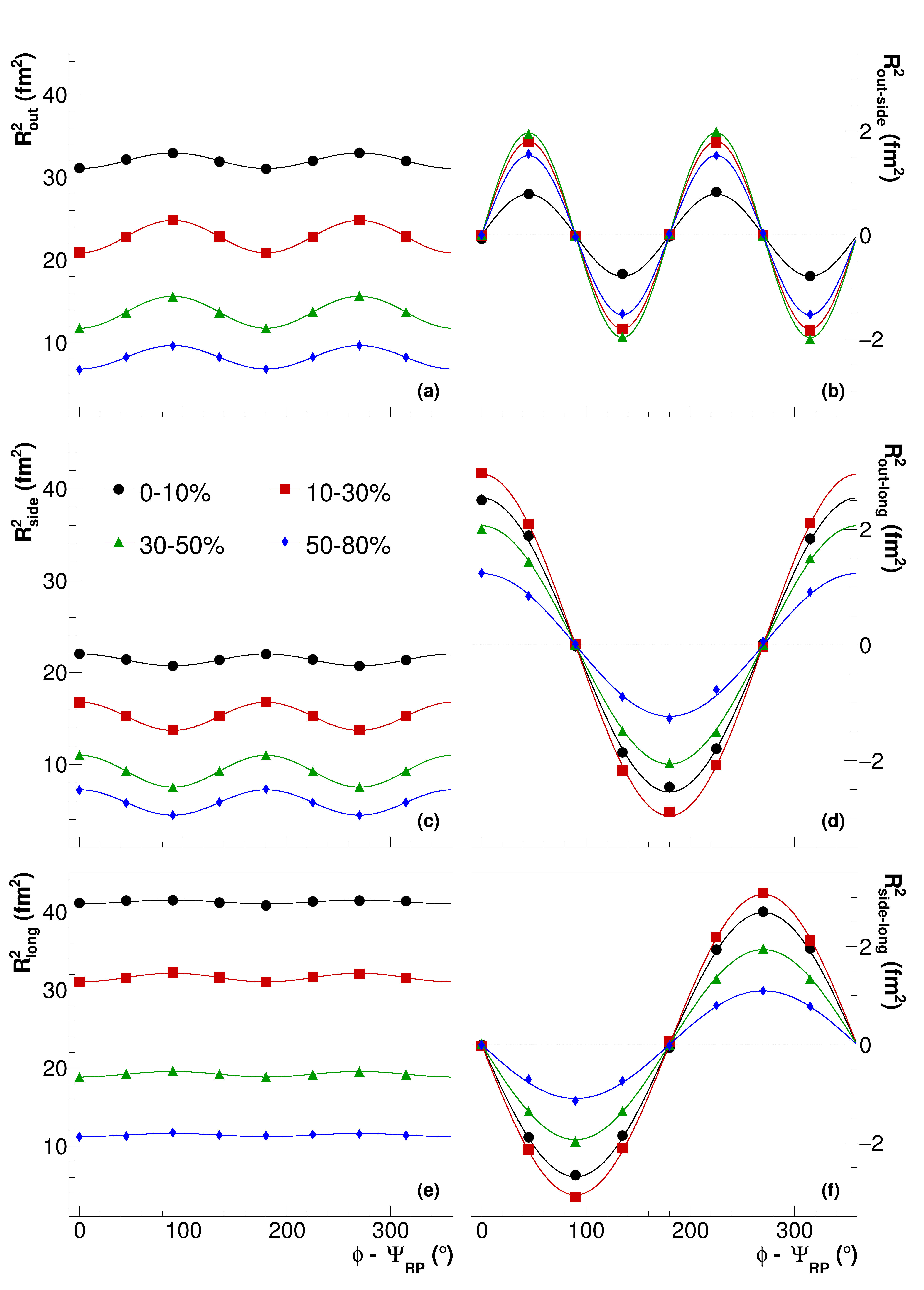}
    \caption{Femtoscopic parameters dependence on the azimuthal angle of the pair of pions with transverse momentum from 0 to 0.1 GeV/c obtained using Au+Au collisions at 7.7 GeV. Different colors corresponds to the  different centralities: black circles - 0-10\%, red squares - 10-30\%, green triangles - 30-50\%, blue diamonds - 50-80\%. Lines correspond to the fit of the femtoscopic parameters oscillations with using Equation~\ref{eq:oscFit}.}
    \label{fig:radiiCent}
\end{figure}

\begin{figure}
    \centering
    \includegraphics[width=1.0\linewidth]{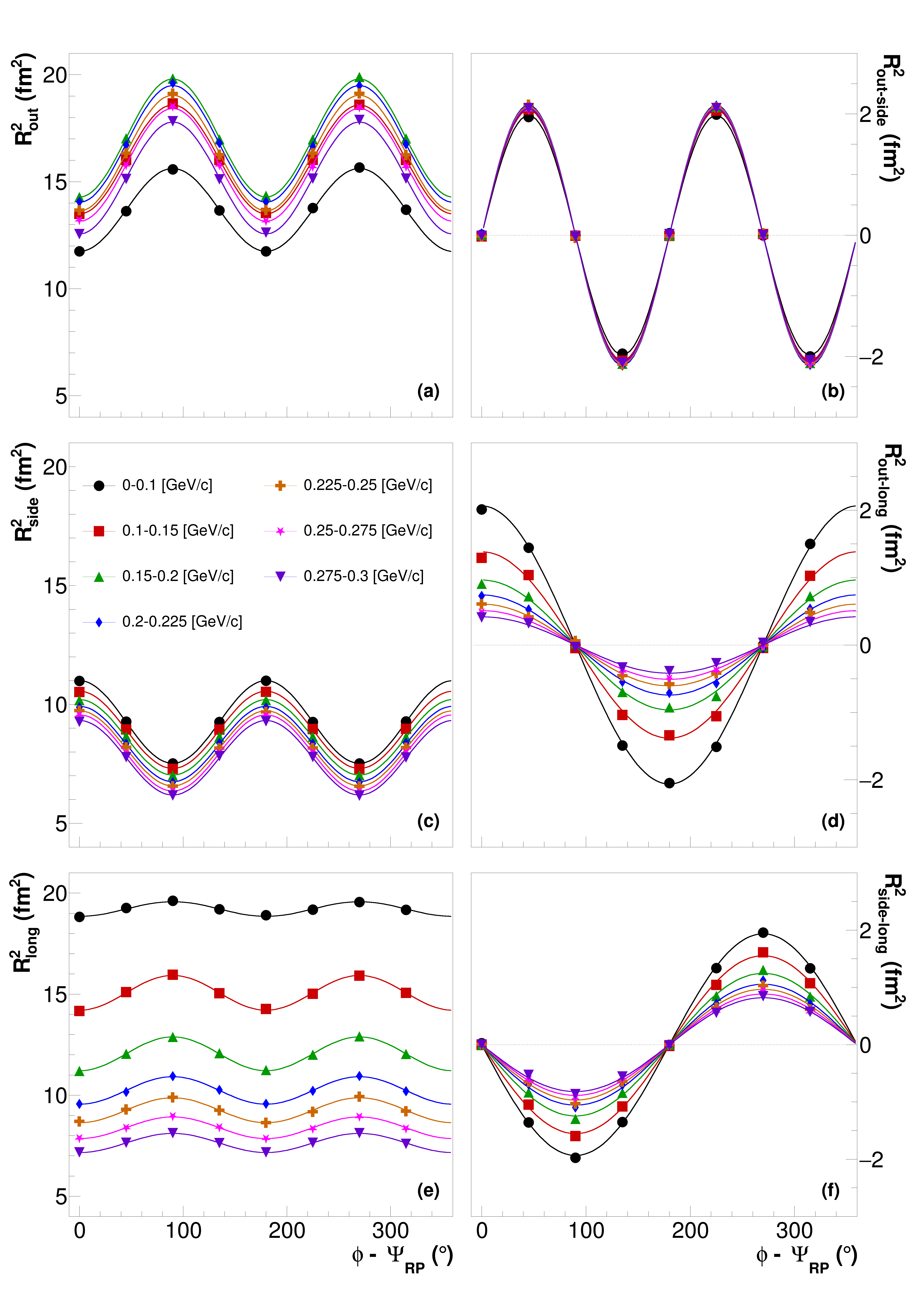}
    \caption{Femtoscopic parameters dependence on the azimuthal angle of the pair of pions produced in collisions with centrality from 30 to 50\% for the Au+Au collisions at 7.7 GeV. Different colors corresponds to the different transverse momentum of the pion pairs. Lines correspond to the fit of the femtoscopic parameters oscillations using Equation~\ref{eq:oscFit}.}
    \label{fig:radiiKt}
\end{figure}

Since reconstructing the tilt angle of the emission source requires azimuthally sensitive analysis, the correlation functions, in addition to being binned by centrality and transverse momentum of the particle pairs, were also divided into eight ranges based on the azimuthal angle of the pion pairs' emission ($\phi$) from the source relative to the reaction plane ($\phi-\Psi_{RP}$). Note that reaction plane ($\Psi_{RP}$) is fixed at zero in the UrQMD. The number of bins for the azimuthal angle division was chosen with future comparison to experimental data in mind. The angles were chosen to capture the maxima of the radii oscillations, as seen in Fig.~\ref{fig:radiiCent} and~\ref{fig:radiiKt}. The lines in Fig. ~\ref{fig:1DCF} correspond to the one-dimensional projections of the three-dimensional fit performed according to the equation~\cite{Wiedemann:1999qn}:
\fontsize{9.5pt}{11pt}\selectfont
\begin{eqnarray}
&C(\textbf{q},\textbf{K},\phi)& \,=  \nonumber \\
&&  1 + \lambda(\textbf{K})\exp\left[- \qquad \sum_{\mathclap{i,j=out,side,long}} \textbf{q}_{i}\textbf{q}_{j}R^{2}_{ij}(\textbf{K},\phi)\right] 
\label{eq:cfFit}
\end{eqnarray}
\normalsize
$\lambda(\textbf{K})$ characterizes the strength of the correlations (or the fraction of coherent particle production)~\cite{Wiedemann:1997cr,STAR:2003nqk,STAR:2004qya}; $\textbf{q}_{i,j}$ is the projected relative pair momentum onto $``out-side-long"$ used in Bertsch-Pratt parameterization~\cite{Pratt:1986cc,Bertsch:1988db}; $R_{ij}$ are the three-dimensional radii and their cross-terms obtained for the $``out-side-long"$ directions. By comparing the one-dimensional projections of the correlation function and its fit, one can observe deviations of the source shape from a Gaussian. These deviations are caused by the impact of the short- and long-lived resonances. A similar effect was observed previously both in models~\cite{Li:2012ta,Kravchenko:2021app} and in  experiment~\cite{STAR:2003nqk,STAR:2004qya}.

Equation~\ref{eq:cfFit} assumes that the freeze-out distribution follows a Gaussian profile in position space, which, in non-central collisions, results in the extracted femtoscopic radii oscillations depending on the azimuthal angle of the pion pair~\cite{Lisa:2000ip}.

It was shown in Ref.~\cite{Heinz:2002au}, that at midrapidity measured femtoscopic radii dependence on $\phi$ can be described by the following Fourier expansions:

\begin{eqnarray}
& & R^2_{out}(\phi)=R^2_{out,0}+2\sum_{n=2,4,6...}R^2_{out,n}\cos(n\phi) \nonumber \\
& & R^2_{side}(\phi)=R^2_{oside,0}+2\sum_{n=2,4,6...}R^2_{side,n}\cos(n\phi) \nonumber \\
& & R^2_{long}(\phi)=R^2_{long,0}+2\sum_{n=2,4,6...}R^2_{long,n}\cos(n\phi) 
\label{eq:oscFit}
\\
& & R^2_{out-side}(\phi)=2\sum_{n=2,4,6...}R^2_{out-side,n}\sin(n\phi) \nonumber \\
& & R^2_{out-long}(\phi)=2\sum_{n=1,3,5...}R^2_{out-long,n}\cos(n\phi) \nonumber \\ 
& & R^2_{side-long}(\phi)=2\sum_{n=1,3,5...}R^2_{side-long,n}\sin(n\phi) \nonumber
\end{eqnarray}

The zero-order Fourier coefficients correspond to the extracted femtoscopic parameters in the azimuthally integrated analysis. Additionally, it is common for Fourier coefficients above the second order to be consistent with zero, a finding that has experimental confirmation~\cite{STAR:2004qya}.

Since the tilt angle of the emission source is largely determined by the collision geometry, it is natural to consider the azimuthal dependence for different centralities. Figure~\ref{fig:radiiCent} shows the oscillations of femtoscopic parameters for four different collision centralities: 0-10\%, 10-30\%, 30-50\%, and 50-80\%. Centrality in this study was determined similarly to the method used in the experiments~\cite{STAR:2008med}.%


Panels $a$, $c$, and $e$ of Figure~\ref{fig:radiiCent} show the oscillations of the standard three-dimensional radii for azimuthally integrated analysis. Panels $b$, $d$, and $f$ display the cross components $``out-side"$, $``out-long"$, and $``side-long"$ which are non-zero only when there is no symmetry with respect to the standard three-dimensional directions. The presence of non-zero values in the cross components of the femtoscopic parameters indicates correlations between the corresponding directions of the relative momentum $\textbf{q}$, which basically are ``tilts" of the correlation function. One can see those in Fig.~\ref{fig:2DCF}. This figure shows the two-dimensional projections of the three-dimensional correlation function for eight ranges of the emission angle of the particle pairs, $\phi$, corresponding to the same cross components as in Fig.~\ref{fig:radiiCent}. For example, the dependence of the correlation function's rotations in the $``out-side"$ projection, shown in the left column of Fig.~\ref{fig:2DCF}, directly leads to the oscillations shown in panel $b$ of Fig.~\ref{fig:radiiCent}.

In our case, cross-terms having non-vanishing values because the analysis is azimuthally sensitive, thus breaking the symmetry with respect to the reaction plane. However, non-zero cross components can also occur in azimuthally integrated analysis, for example, in the absence of boost invariance in the collision system~\cite{Wiedemann:1999qn} or simply when considering the system at forward/backward instead of mid-rapidity. It was shown that $R_{out-long}$, for example, deviates from zero in fixed-target experiment~\cite{NA49:1998syj} due to this reasoning.

In Fig.~\ref{fig:radiiCent}, the lines represent fits according to Equations~\ref{eq:oscFit}. The extracted Fourier coefficients can be used to calculate the tilt of the homogeneity region in space~\cite{Lisa:2000ip, Mount:2010ey}:
\begin{equation}
\theta_{out-long} = \frac{1}{2}\tan^{-1}\left( \frac{4R^2_{out-long,1}}{R^{2}_{long,0}-R^{2}_{side,0}+2R^{2}_{side,2}} \right)
\label{eq:tiltOL}
\end{equation}
or
\begin{equation}
\theta_{side-long} = \frac{1}{2}\tan^{-1}\left( \frac{-4R^2_{side-long,1}}{R^{2}_{long,0}-R^{2}_{side,0}+2R^{2}_{side,2}} \right)
\label{eq:tiltSl}
\end{equation}
where the number below indicates the order of the Fourier coefficient. Additionally, a small correction in the calculation will be made due to the fact that the measurement of femtoscopic parameters as a function of $\phi$ cannot be performed as a continuous function. Thus, the obtained values need to be adjusted according to~\cite{Heinz:2002au, Mount:2010ey}:
\begin{equation}
R^{'2}_{\mu,n}=\frac{n\Delta\phi/2}{\sin(n\Delta\phi/2)}R^2_{\mu,n}
\label{eq:binWidth}
\end{equation}
where $R_{\mu,n}$ are obtained radii for corresponding Furier coefficient from Eq.~\ref{eq:oscFit} and $\Delta\phi$ is the size of the bin on  the azimuthal
angle of the pair of pions.

As previously mentioned, it is quite natural to assume that the rotation of the system in space will vary somewhat depending on the centrality of the collision. On the other hand oscillations dependence on $k_{T}$ shown on the Figure~\ref{fig:radiiKt} are nontrivial. This figure shows the same data as Figure~\ref{fig:radiiCent}, but for different transverse momenta of the pion pairs. Panels $b$, $d$, and $f$, as before, show the cross components. It can be seen that in the $``out-long"$ and $``side-long"$ components, there are noticeable differences in the oscillations obtained for different pair momenta. As shown in Eq.~\ref{eq:tiltOL} and Eq.~\ref{eq:tiltSl}, these components play crucial role for calculating rotation the homogeneity region.

\begin{figure}
    \centering
    \includegraphics[width=1.0\linewidth]{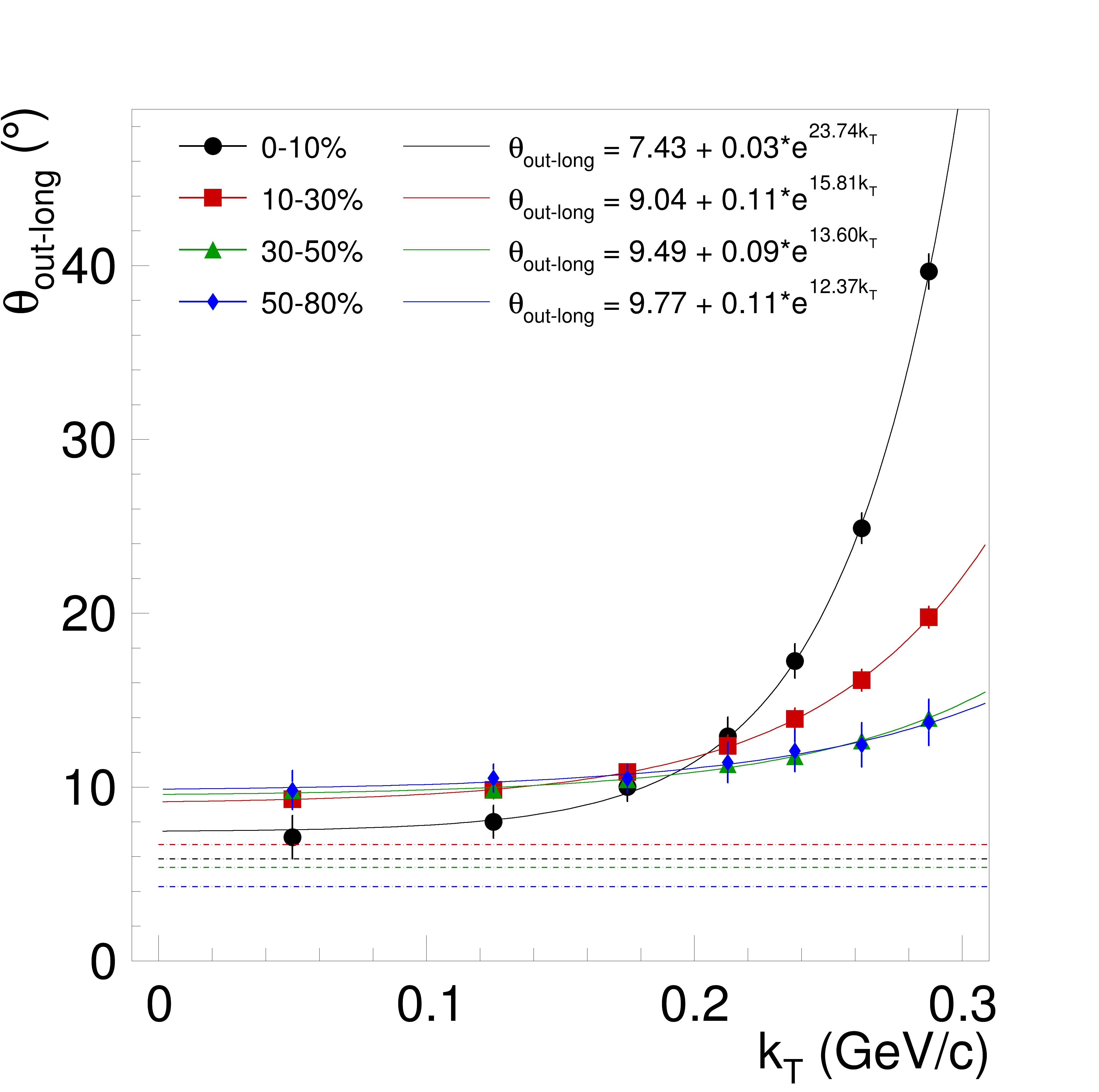}
    \caption{Dependence of homogeneity region tilt according to the Eqn.~\ref{eq:tiltOL} on the transverse momentum of pion pairs for multiple collision centralities at 7.7~GeV. The lines represent fits with an exponential function. }
    \label{fig:ol_7.7}
\end{figure}

\begin{figure}
    \centering
    \includegraphics[width=1.0\linewidth]{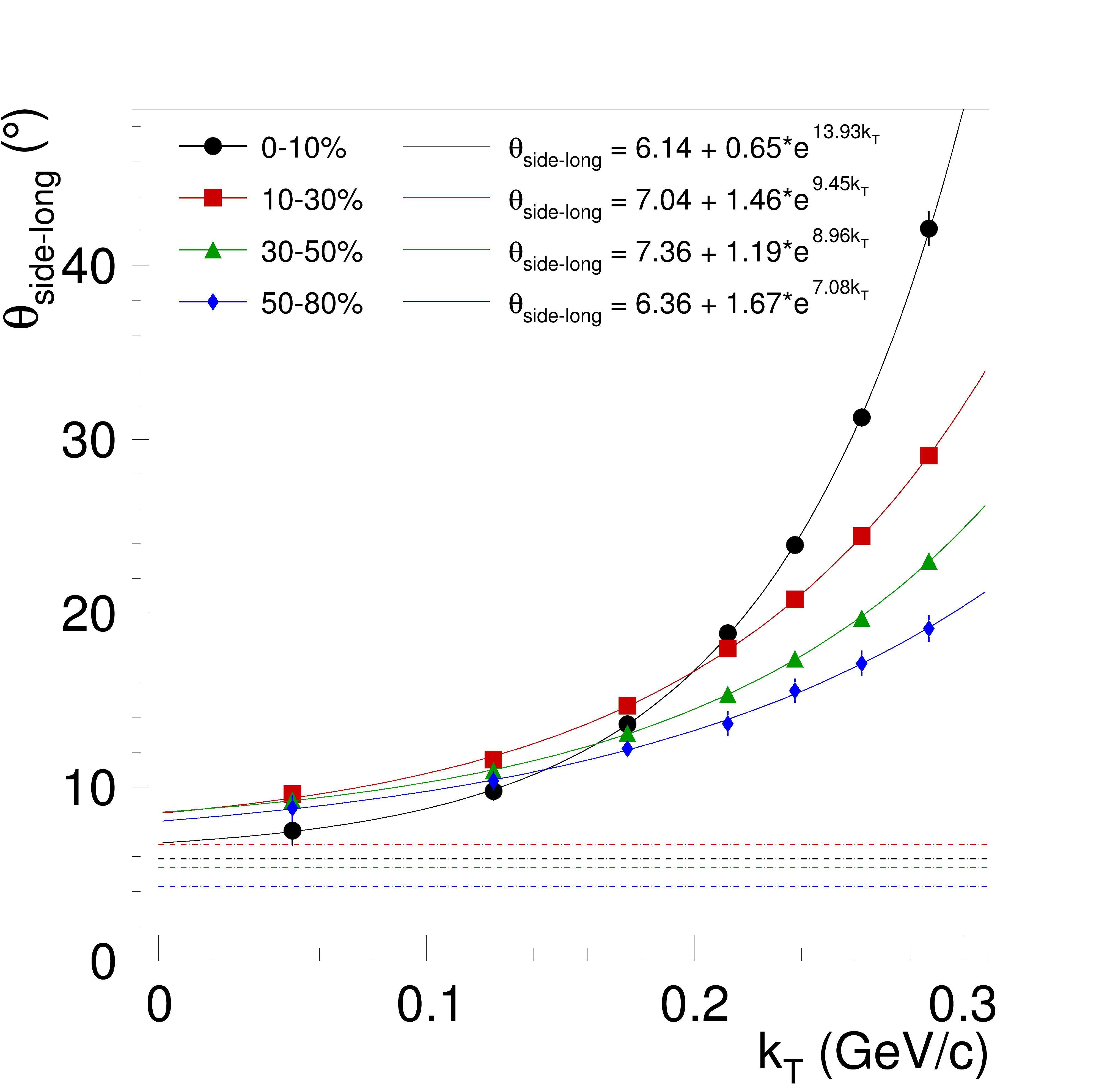}
    \caption{Dependence of homogeneity region tilt according to the Eqn.~\ref{eq:tiltSl} on the transverse momentum of pion pairs for multiple collision centralities at 7.7~GeV. The lines represent fits with an exponential function.}
    \label{fig:sl_7.7}
\end{figure}

\begin{figure}
    \centering
    \includegraphics[width=1.0\linewidth]{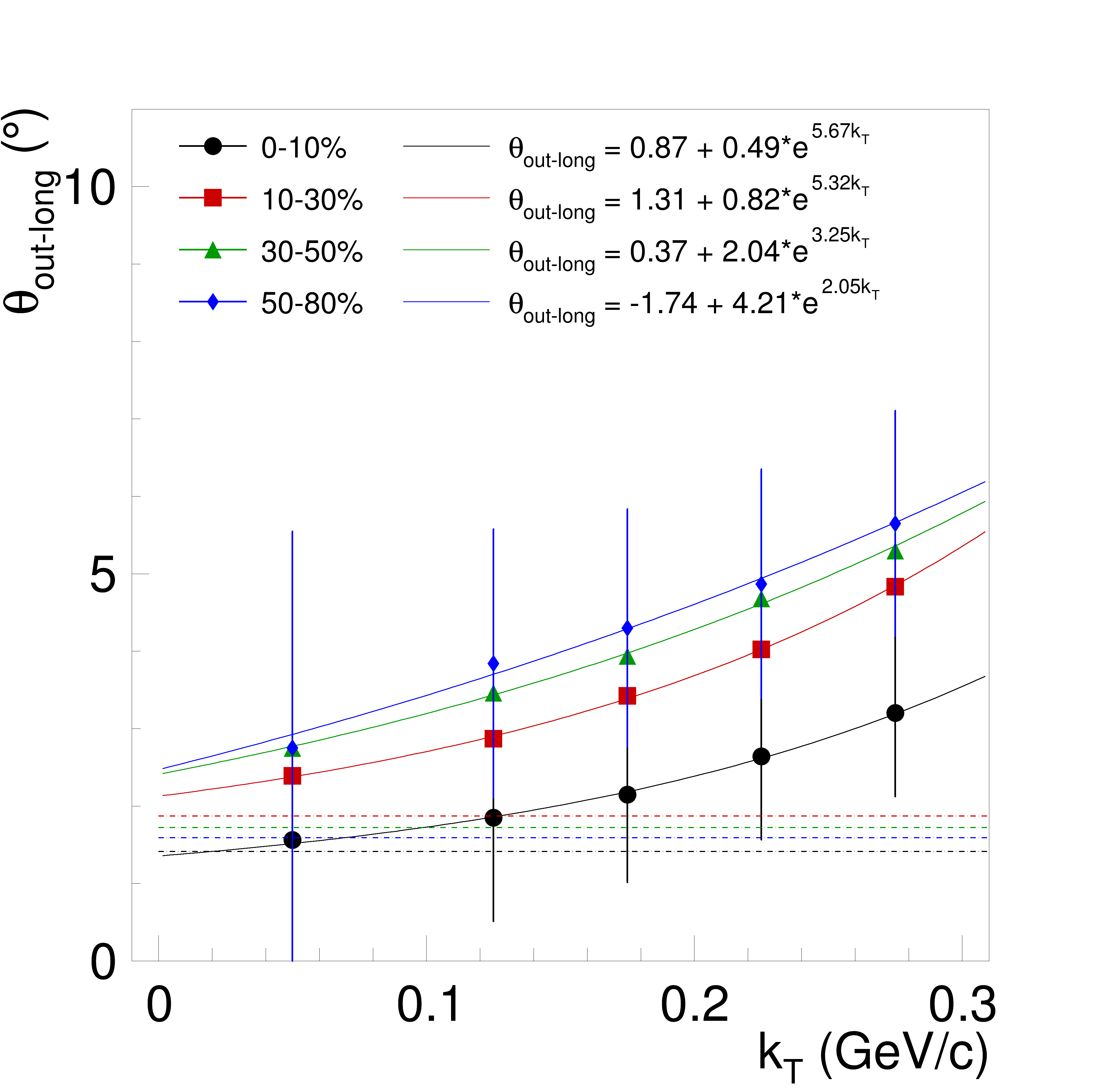}
    \caption{Same as Figure~\ref{fig:ol_7.7}, but for 27 GeV.}
    \label{fig:ol_27}
\end{figure}

\begin{figure}
    \centering
    \includegraphics[width=1.0\linewidth]{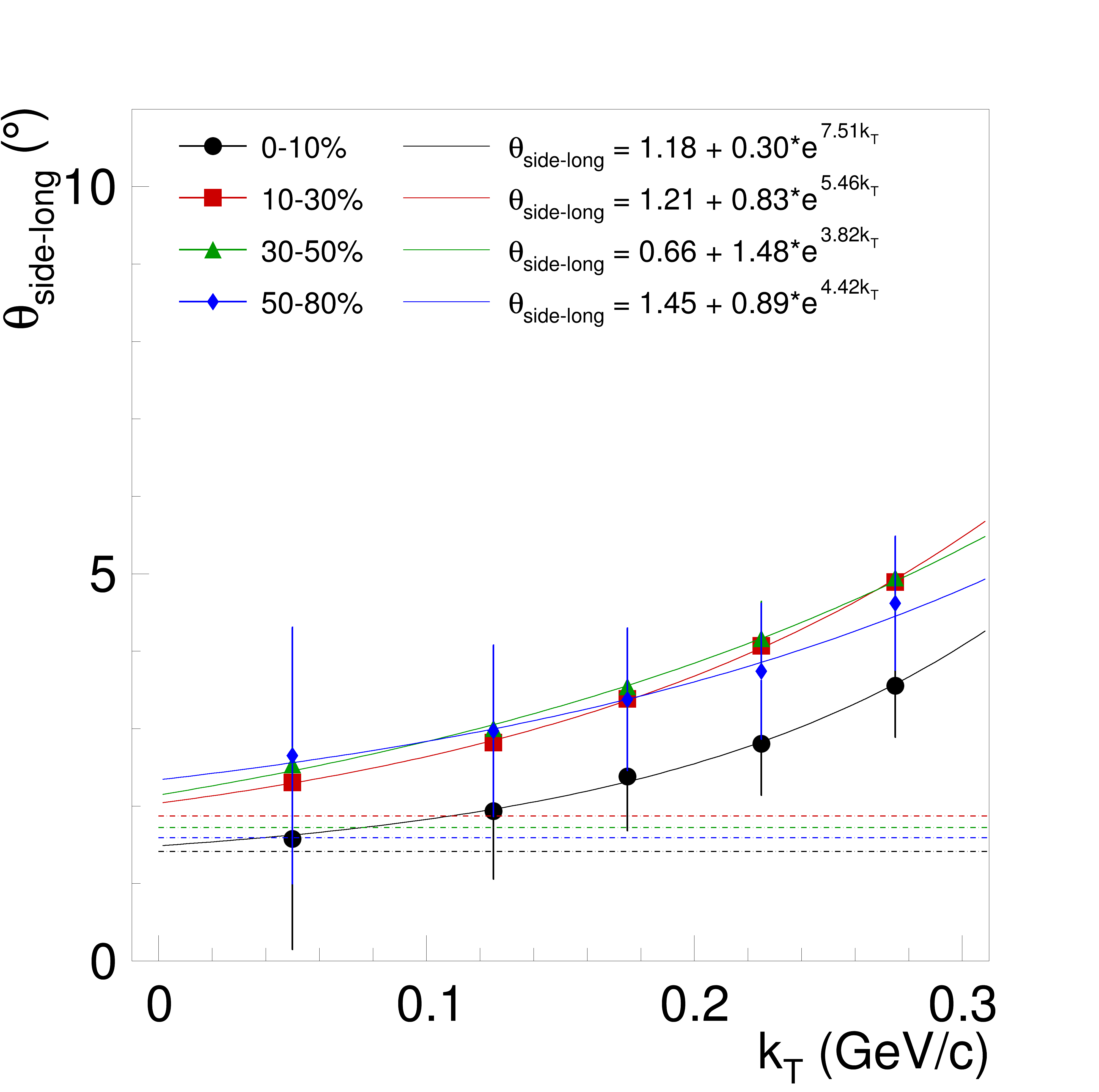}
    \caption{Same as Figure~\ref{fig:sl_7.7}, but for 27 GeV.}
    \label{fig:sl_27}
\end{figure}

Figures~\ref{fig:ol_7.7} and~\ref{fig:sl_7.7} show the dependence of the homogeneity regions tilt on the transverse momentum of the particle pair, calculated using Eq.~\ref{eq:tiltOL} and~\ref{eq:tiltSl}, respectively. Both equations should generally reflect the same rotation angle of the system in space; however, as can be seen from Fig.~\ref{fig:sl_7.7} and~\ref{fig:ol_7.7}, the angles can be quite different depending on $k_{T}$. This difference may be due to the fact that by definition the $``out"$ component of the Bertsch-Pratt parameterization is directed in the same way as the $k_{T}$ of the pion pair. Therefore, the difference between $\theta_{side-long}$ and $\theta_{out-long}$ may be due to the developing system's dynamics. It is interesting to note that at the smallest considered pion pair momentum, the values of $\theta_{side-long}$ and $\theta_{out-long}$ coincide within the statistical uncertainties.

It can also be seen that the dependence on the transverse angle of the pion pair is quite strong, ranging from approximately 5 degrees (at the lowest considered $k_{T}$) to around 40 degrees (at the highest $k_{T}$ in this study). The existence of this dependence is driven by the fact that using femtoscopy method we are measuring not the tilt of the whole source of the pions but rather the tilt of homogeneity regions~\cite{Akkelin:1995gh}. 
As shown earlier, the larger the $k_{T}$ range we consider, the smaller the corresponding regions of homogeneity~\cite{Lisa:2005dd}. 
Thus, the larger the $k_{T}$ we consider, the less correspondence to the rotation of the entire system. This fact is very important when measuring the system's rotation angle in an experiment, as commonly used detectors have lower limits on the momentum of particles they can measure. For example, the effectively reconstructed momentum registered by the Time Projection Chamber in the STAR experiment starts at $\sim$0.2 GeV/c~\cite{Anderson:2003ur}. As seen from Fig.~\ref{fig:ol_7.7} and~\ref{fig:sl_7.7}, at this momentum, the system's rotation begins to deviate significantly from what was obtained at the lowest momenta. It is also important to remember that experimental measurements are statistically limited, and the $k_{T}$ spectrum at these values has orders of magnitude fewer entries than, for example, at $k_{T}$ = 0.25 GeV/c. Therefore, measurements at very low $k_{T}$ in experiments will be quite challenging.

With this fact in mind we have made approximation of the $k_{T}$ dependence down to zero value using the most suitable function. Lines on Fig.~\ref{fig:ol_7.7} and~\ref{fig:sl_7.7} show the exponential approximation of dependencies for different centralities. The parameters of the approximations are also shown on the figures. The dashed lines there indicate the rotation values of the system obtained from the fit of the three-dimensional distribution of freeze-out coordinates using Eq.~\ref{eq:freeze-out}. As one can see, the approximated $k_{T}$ dependencies at low momenta are quite close to the values obtained from the fit of the pion freeze-out distribution. It is important to remember that both the femtoscopic analysis and the freeze-out coordinate analysis assume Gaussian sources. However, as discussed earlier, both exhibit a small degree of non-Gaussian behaviour, which may affect the final results obtained here.

The dependence of the tilt angle on the transverse momentum of a pair of particles itself may also be of interest. One can think about the $k_{T}$ dependence of the tilted homogeneity regions in order to describe directed flow of heavier particles. We are leaving for future studies detailed investigation of this phenomenon.

\section{Summary}

Many models, in their attempts to describe forward-backward moving systems, assume the presence of boost invariance in the resulting particle emission source. However, as multiple measurements of directed flow as a function of rapidity show, the emerging system dynamics are never truly boost invariant, even at LHC energies~\cite{ALICE:2013xri}. Thus it is important to take into account the tilt of the emitting source away from the beam direction.

In this paper, we investigated the tilt angle of the pion emission source, which influences the system's development after kinetic freeze-out. We extracted the tilt angle both directly from the spatial freeze-out coordinate distribution of pions and using the method of azimuthally sensitive femtoscopy. It was shown that the transverse momentum of the particle pair is a crucial factor when extracting the tilt angle parameter from asHBT analysis.

In experiments, access to very low momenta particles is not always feasible due to detector acceptance. However, in this study it was demonstrated that the dependence of the tilt angle parameter of the emission source follows an exponential trend, and the extrapolated value at $k_{T}$~=~0 GeV/c is quite close to the tilt angle directly extracted from the three-dimensional freeze-out distribution of particles. The remaining discrepancy may be attributed to the well-known deviation of the applied models from Gaussianity, as seen in Fig.~\ref{fig:freeze-out} and~\ref{fig:1DCF}.

The study of how the tilt angle of homogeneity regions depends on the transverse momentum of particles can be intriguing on its own without extrapolation to zero.

Instead of attributing the directed flow to the tilt of the entire emission source~\cite{Du:2022yok,Bozek:2010bi,STAR:2017ykf,Chatterjee:2017ahy,STAR:2019clv}, 

it can be more effective to consider the regions of homogeneity as the underlying cause. By accurately determining the tilt angles of these regions for different momenta or types of particles, we can achieve a more precise description of the directed flow across various particles and momenta. This approach offers possibly clearer and more detailed understanding of the directed flow 
phenomena.

\begin{acknowledgments}
This work supported by The U.S. Department of Energy Grant DE-SC0020651.
\end{acknowledgments}

\bibliography{apssamp}

\end{document}